\documentclass[conference]{IEEEtran}
\IEEEoverridecommandlockouts
\usepackage{cite}
\usepackage{amsmath,amssymb,amsfonts}
\usepackage{algorithmic}
\usepackage{graphicx}
\usepackage{tabularx}  
\usepackage{textcomp}
\usepackage{array}
\usepackage{xcolor}
\usepackage[utf8]{inputenc}
\usepackage{tcolorbox}
\usepackage{colortbl}

\def\BibTeX{{\rm B\kern-.05em{\sc i\kern-.025em b}\kern-.08em
    T\kern-.1667em\lower.7ex\hbox{E}\kern-.125emX}}
\begin{document}

\title{Identifying Factors Contributing to ``Bad Days'' for Software Developers: A Mixed-Methods Study\\

}

\makeatletter
\newcommand{\linebreakand}{%
  \end{@IEEEauthorhalign}
  \hfill\mbox{}\par
  \mbox{}\hfill\begin{@IEEEauthorhalign}
}
\makeatother

\author{
  \IEEEauthorblockN{Ike Obi}
  \IEEEauthorblockA{
    \textit{Purdue University}\\
}
  \and
  \IEEEauthorblockN{Jenna Butler}
  \IEEEauthorblockA{
    \textit{Microsoft Corporation}\\
}
  \and
  \IEEEauthorblockN{Sankeerti Haniyur}
  \IEEEauthorblockA{
    \textit{Microsoft Corporation}\\
}
  \linebreakand 
    \IEEEauthorblockN{Brian Hassan}
  \IEEEauthorblockA{
    \textit{Microsoft Corporation}\\
}
  \and
  \IEEEauthorblockN{Margaret-Anne Storey}
  \IEEEauthorblockA{
    \textit{University of Victoria}\\
}
  \and
  \IEEEauthorblockN{Brendan Murphy}
  \IEEEauthorblockA{
    \textit{Microsoft Corporation}\\
}
}

\maketitle

\vskip -2cm  

\begin{abstract}
Software development is a dynamic activity that requires engineers to work effectively with tools, processes, and collaborative teams. As a result, the presence of friction can significantly hinder productivity, increase frustration, and contribute to low morale among developers. By contrast, higher satisfaction levels are positively correlated with higher levels of perceived productivity. Hence, understanding the factors that cause bad experiences for developers is critical for fostering a positive and productive engineering environment. In this research, we employed a mixed-method approach, including interviews, surveys, diary studies, and analysis of developer telemetry data to uncover and triangulate common factors that cause ``bad days'' for developers. The interviews involved 22 developers across different levels and roles. The survey captured the perception of 214 developers about factors that cause them to have ``bad days'', their frequency, and their impact on job satisfaction. The daily diary study engaged 79 developers for 30 days to document factors that caused ``bad days'' in the moment. We examined the telemetry signals of 131 consenting participants to validate the impact of bad developer experience using system data. Findings from our research revealed factors that cause ``bad days'' for developers and significantly impact their work and well-being. We discuss the implications of these findings and suggest future work. 

\end{abstract}


\section{Introduction}

Enhancing developer experience \cite{fagerholm2012developer} is crucial for improving software productivity for several reasons. A positive developer experience reduces unnecessary blockers and enables developers to concentrate on their core responsibilities of writing, testing, and deploying code rather than being held by back tools and processes \cite{forsgren2024devex}. Prior research has also provided further reasons for the importance of improving developer experience. For instance, Forsgren et al. \cite{Forsgren_2024} highlighted that developers who had more time for deep work demonstrated a remarkable 50\% increase in productivity compared to their counterparts with less uninterrupted time. Recent work by Noda et al. \cite{noda2023devex} and Srivastava et al. \cite{srivastava2020developer} also showed that companies that provided suitable work environments for developers achieved revenue growth that was five times greater than their competitors with suboptimal work environments. In the same vein, prior research by Doerrfeld \cite{Doerrfeld_2022} showed that developer experience plays a crucial role in talent retention, with 63\% of developers in their study mentioning that they consider their experience at the workplace as a key factor when deciding to remain at or leave the organization.  

Building upon these foundational insights, recent work by other scholars explored different ways of improving developer experience. This includes studying what constitutes a good day for developers \cite{meyer2019today}, investigating the impact of flow state on developer satisfaction and productivity \cite{muller2015stuck}, and examining the intersection between developer unhappiness, productivity, and developer attrition rates through a systematic analysis \cite{graziotin2019happiness}. Other research has studied how specific tools and processes like pull requests and build times contribute to negative experiences for developers \cite{egelman2020predicting}. Despite these valuable contributions, as of yet, very limited work has engaged with developers to holistically investigate from their perspective the factors that cause them to have a \textbf{``bad day''}, its impact on their productivity and well-being, and their suggestions on how to mitigate those experiences. Even more so, very limited studies have paired developer feedback with telemetry data to examine and validate developer concerns about poor developer experiences. We do not offer a definition of what a ``bad day'' means as we are interested in using the subjective term of a ``bad day''\footnote{We also find through our research that what a bad day means to developers varies, and even varies for a developer from one day to the other.} to \emph{elicit} the common problems that make developers feel they are having a bad day at work. Particularly, we are focused on being able to detect those problems and drive them down as a means of improving the developer experience.

In this research, we employed a mixed-methods approach, including interviews, surveys, diary study, and telemetry data analysis to investigate and validate factors that cause ``bad days'' for developers. Our study engaged 22 software developers in interview sessions, 214 developers participated in our survey study, and 79 developers participated in a diary study. Results from these three approaches were then validated using the telemetry data of 131 consenting developers to examine the difference between the log data of those who reported having a ``bad day'' versus those who did not report having a ``bad day''. By combining these approaches, we sought to uncover, triangulate, and examine factors that cause developers to have a ``bad day'' from multiple viewpoints and validate the measurable aspects of the findings using telemetry data. These results allowed us to answer our research questions, including:

\begin{enumerate}
    \item \textbf{RQ1:} What factors cause a ``bad day'' for developers?
    \item \textbf{RQ2:} In what ways do developers describe how these ``bad day'' factors impact them and their work?
    \item \textbf{RQ3:} Can we observe ``bad day'' factors using telemetry data?
\end{enumerate}

Our research makes the following contributions: 1) A comprehensive view of factors that cause ``bad day'' for developers, including both technical and non-technical factors as both factors can lead to a poor developer experience, 2) Insights into the impact of negative developer experiences on developer satisfaction and well-being, thus providing a better understanding of this phenomenon and how developers could be supported, 3) Validation through telemetry analysis, the concrete impact of factors such as build times and pull request processes on developer satisfaction, establishing quantifiable thresholds that correlate with negative experiences. Overall, through this research, we contribute new insights into the concept of ``bad developer days'' beyond what is available in current literature.

The rest of this paper is organized as follows. In the following section, we build on prior work to motivate the contributions of our study, next, we provide a detailed description of our methods, followed by our findings and the discussion of results from this research, and then conclude by highlighting the implications of our findings.

\section{Related Work}

In this section, we highlight prior research that has discussed the connection between the emotional states of developers and productivity and well-being levels. We also highlight prior work that has explored approaches for capturing and measuring developer productivity and challenges and build on those works for the contributions we make in this research.

\subsection{The Role of Emotions in Developer Productivity}

Bad work days often lead to negative emotions and low developer productivity. Several scholars have investigated the connection between developer emotions and productivity, seeking to foreground how the negative or positive emotional state of developers influences their work outcomes \cite{storey2019towards,razzaq2024systematic,meyer2017work,girardi2021emotions}. Graziotin et al. \cite{graziotin2014happy} investigated the relationship between developer affective states (emotions and moods), creativity, and analytical problem-solving skills. They engaged 42 computer science students in a study to measure how their emotional states correlated with performance on problem-solving tasks. Their research found that happy software developers are better problem solvers, at least in terms of analytical abilities. Their finding suggests that fostering positive emotional states among developers may lead to improved productivity. Müller and Fritz \cite{muller2015stuck} also investigated the developer's emotional states to better understand how emotions such as frustrations and happiness correlate with the performance of developers and their ability to focus in a flow state. Their research revealed that positive emotions enhance the flow state of developers and, thus, the ability of developers to do good work.

Meyer et al. \cite{meyer2019today} also explored what makes a developer's day good. Their study collected self-reported data from developers over four months, analyzing how developers spent their time and what factors contributed to their perception of a good workday. Findings from their research revealed that the satisfaction and productivity of developers are significantly influenced by their ability to control their workday and minimize unplanned disruptions. Along a similar vein, Graziotin et al. \cite{graziotin2017unhappy} explored the consequences of unhappiness among software developers while developing software. They conducted a qualitative analysis of survey responses from 181 developers to identify and categorize the negative effects of unhappiness on developers themselves, the software development process, and the resulting products they create. From their study, they found the highest impact of unhappiness is on productivity and performance. Their findings also showed that unhappiness leads to mental health issues among developers through issues like low self-esteem, high anxiety, burnout, stress, and potentially more serious disorders like depression.

\subsection{Measuring and Understanding Developer Productivity}

Several scholars have explored approaches for characterizing, measuring, and predicting the productivity levels and fluctuations of developers \cite{d2024measuring}. Forsgren et al. \cite{forsgren2021space} introduced the SPACE framework, a comprehensive model designed to measure and understand developer productivity. Implicit in their framework is that developer productivity is complex and multi-faceted and requires a balanced and comprehensive approach to measure and capture its dimensions. Brown et al. \cite{brown2023using} developed a logs-based metric to identify periods of focused work and flow in software engineers. Their study employed a mixed-methods approach, including diary studies and longitudinal surveys, to validate the metric against self-reported data. Findings from their research revealed that data from the logs-based metric correlated significantly with self-reported experiences of focus and flow showing the promise of this approach.

Egelman et al. \cite{egelman2020predicting} also investigated the negative interpersonal experiences, termed "pushback," developers encounter during code reviews. They surveyed 1,317 developers through a mixed-methods approach and analyzed code review logs to identify the prevalence and predictors of negative feelings associated with code reviews. They found 57\% of developers reported experiencing negative feelings related to code reviews at least once a quarter, with 26\% experiencing such feelings monthly. This showed that although pushback during code reviews is relatively rare, it still has a significant negative impact on developer satisfaction when it occurs. Meyer et al. \cite{meyer2014software} investigated how software developers perceive and assess their own productivity. They conducted two studies - a survey of 379 professional developers and an observational study of 11 developers - to understand how developers perceive productive vs. unproductive activities and how they measure productivity. Their study showed that developers perceive their days as productive when they complete many tasks or large tasks without significant interruptions. Coding was seen as the most productive activity, while meetings were viewed as both productive and unproductive, depending on their nature. Murphy-Hill et al. \cite{murphy2019predicts} investigated factors that predict productivity for developers by surveying 622 developers across three companies. Their survey included questions about self-rated productivity, productivity factors, and demographic variables. Their study revealed both technical and non-technical factors which improve productivity.

From a more practical viewpoint, research by Lu et al. \cite{lu2008learning} highlights that flaky tests not only disrupt the continuous integration process but also increase debugging efforts, thereby diminishing developer productivity. Similarly, prior work by Ko et al. \cite{ko2007information}, highlights that inadequate documentation negatively impacts developers in different ways, including by increasing the learning curve for new tools, impeding troubleshooting and maintenance tasks, and ultimately leading to frustration and reduced productivity among developers.  Wang et al. \cite{wang2013empirical} also pointed out that slow build times are a critical bottleneck that leads to poor developer concentration and workflow efficiency. Their findings suggest a direct correlation between faster build processes and improved developer satisfaction and productivity.

\section{Methods}
In this study, we employed a mixed methods approach to investigate the factors that cause ``bad days'' for developers. Specifically, our approach paired qualitative methods, including interviews and diary studies, with quantitative methods including survey studies and telemetry data analysis to elicit and measure factors that cause ``bad days'' for developers from different viewpoints. 

\begin{figure}[h!]
    \centering
    \includegraphics[width=0.5\textwidth, trim=0cm 3cm 0cm 3cm, clip]{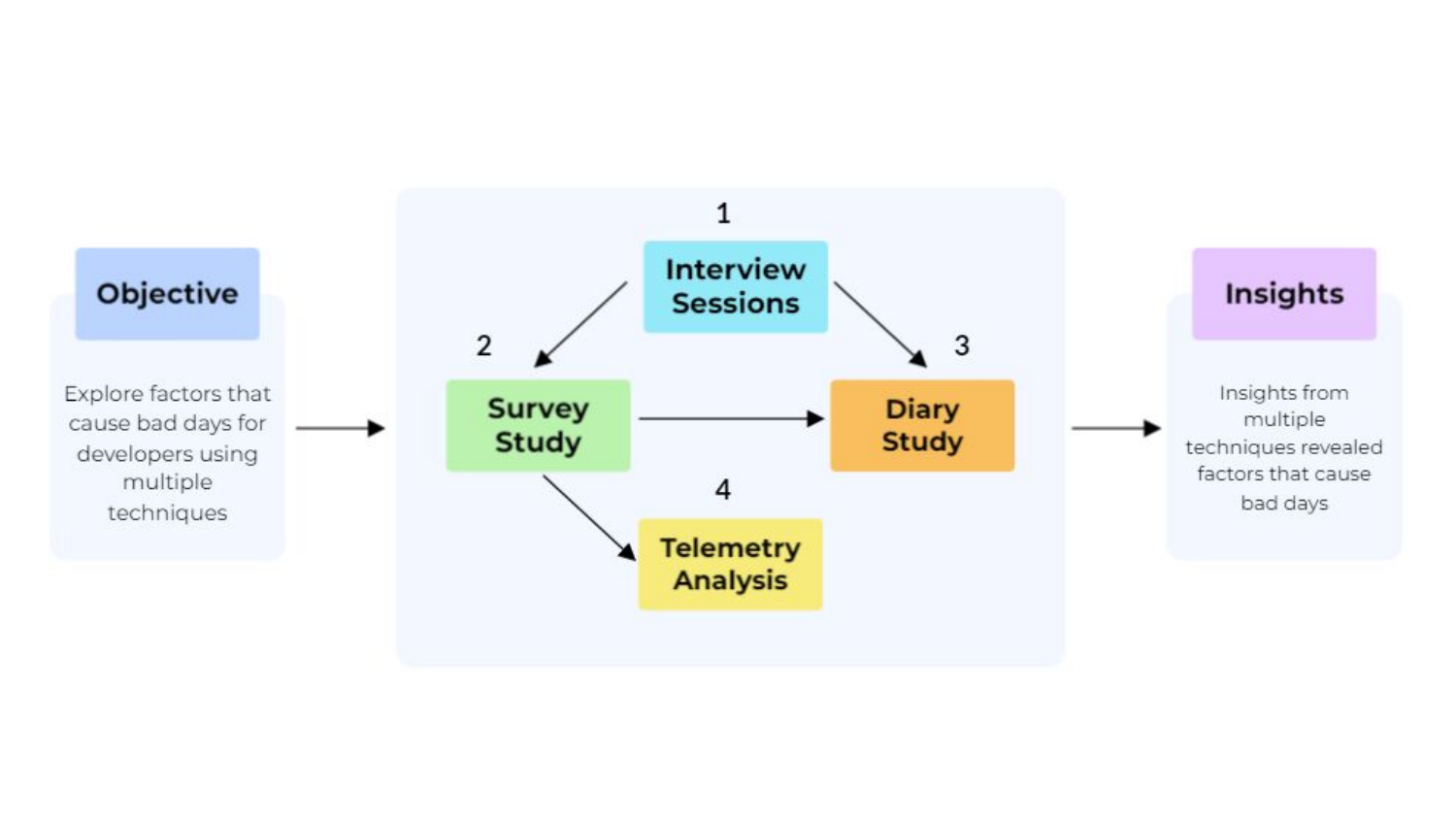}
    \caption{Our research process involved a mixed-method approach that allowed us to foreground and triangulate insights from different approaches}
    \label{fig:researchprocess}
\end{figure}

\subsection{\textbf{Data Collection}}
Across our methods, we had varying levels of participation from developers. Figure \ref{fig:datacollection} shows the number of participants in each study method, and the following sections will go into more detail about the methodology and data collection for each.

\begin{figure}[ht]
    \centering
    \includegraphics[width=0.4\textwidth]{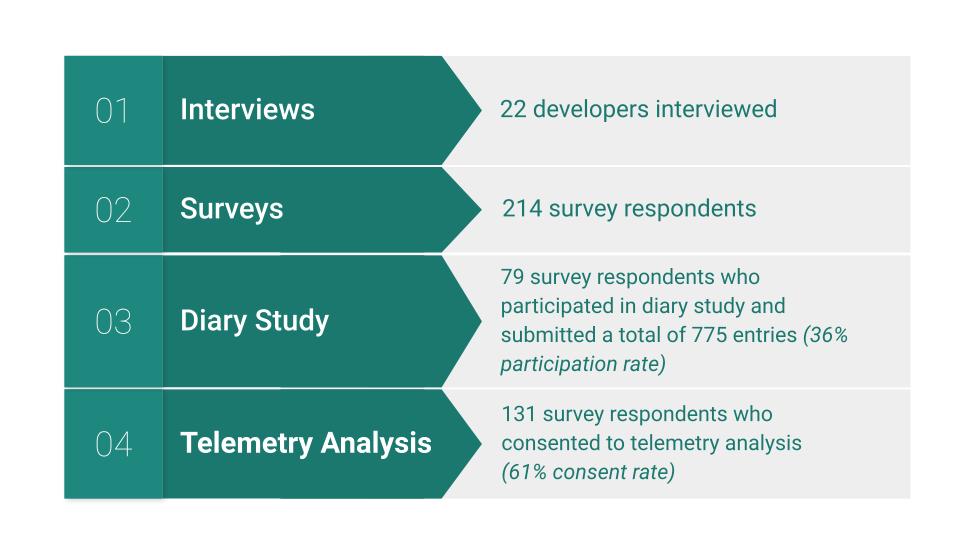}
    \caption{Number of participants in each of our mixed-method approaches}
    \label{fig:datacollection}
    \vspace{-15pt}
\end{figure}

\subsubsection{Interviews}
We interviewed 22 software developers to understand from their perceptive, the factors that are causing them to have ``bad days'' in their development work. We recruited developers with diverse backgrounds, roles, and experience levels to ensure we capture the perspectives of a broad range of developers within our organization. Our recruitment employed snowball sampling was conducted through internal company media that involved posters, word-of-mouth, and email outreach. The recruitment message outlined the purpose of the study, which was to investigate the everyday work experiences of software developers, with a particular focus on challenges and frustrations that occur during the workday that cause them to have a ``bad day''. The inclusion criteria for participating in the study required that participants be software developers or engineers within the organization. Hence, non-engineers were not included in the study.

We collected the interview data using a semi-structured interview protocol that was prepared ahead of time. Each interview lasted approximately 30-60 minutes and was conducted via a video conferencing platform. All interviews were audio-recorded with the consent of the participants and later transcribed for analysis. We ensured to recruit across levels, role types, and locations -- Table \ref{table:interviewprofiles} lists the demographic spread of the interviewees. We had a good distribution of interviewees across Level bands. Although we had an over-representation of Individual Contributors and interviewees in the USA, this is distribution representative of the developer population at the company. 

\vspace{-7pt}

\begin{table}[h!]
\renewcommand{\arraystretch}{1.2} 
\caption{Demographic Profile of Interviewees}
\centering
\begin{tabular}{|l|l|l|} \hline
\rowcolor{gray!30}
\textbf{Demographic} & \textbf{N} & \textbf{\%} \\ \hline
\rowcolor{gray!10}
\multicolumn{3}{|l|}{\textbf{Level}} \\ \hline
\hspace{1em} Software Developer I & 7 & 32\% \\ \hline
\hspace{1em} Software Developer II & 6 & 27\% \\ \hline
\hspace{1em} Senior Software Developer & 5 & 23\% \\ \hline
\hspace{1em} Principal Software Developer & 4 & 18\% \\ \hline
\rowcolor{gray!10}
\multicolumn{3}{|l|}{\textbf{Role Type}} \\ \hline
\hspace{1em} Individual Contributor & 21 & 95\% \\ \hline
\hspace{1em} People Manager & 1 & 5\% \\ \hline
\rowcolor{gray!10}
\multicolumn{3}{|l|}{\textbf{Location}} \\ \hline
\hspace{1em} USA & 15 & 68\% \\ \hline
\hspace{1em} Canada & 3 & 14\% \\ \hline
\hspace{1em} Ireland & 2 & 9\% \\ \hline
\hspace{1em} India & 1 & 5\% \\ \hline
\hspace{1em} New Zealand & 1 & 5\% \\ \hline
\end{tabular}
\label{table:interviewprofiles}
\end{table}

During the interview, we explored four main areas with the developers, including asking them to define what they understand by the term ``bad day for developers'' and to further describe what constitutes a ``bad day'' at work, including associated thoughts and feelings, and how these differ from typical challenging days. The developers were also prompted to share recent examples of ``bad days,'' discussing the frequency, factors, and events that contributed to these experiences. They were also asked about patterns or recurring situations that tend to trigger such days. In the third section of the interview, we focused on eliciting from the participants how ``bad days'' affect productivity, work performance, emotions, well-being, and its impact on their interactions with their teammates and colleagues. During the final stage of the interview, developers were asked about coping strategies, helpful resources, and potential changes to their work environment that could prevent or mitigate ``bad days.'' Overall, through these interviews, we generated qualitative data on the perception of developers within our organization about factors that make a developer's day go bad at work.

\subsubsection{Survey}
Following the interviews, we conducted a survey study with 214 developers to further investigate the factors that are causing them to experience a ``bad day'' from a wider audience. We aimed to find common themes from the interviews and surveys that reflect developer sentiment within the organization. Participants were recruited through an organization-wide email inviting them to take part in the study. The email included an informed consent form outlining the purpose of the study, confidentiality measures around their privacy, and how their data will be handled at the end of the study. Participants were then directed to an online survey platform where they took the survey. This study was approved by our organization's ethics review board and followed best practices to protect the privacy and well-being of the developers who participated in the study. Only members of the research team who are all trained researchers have access to the data from this study.

The survey consisted of four sections. We wanted the survey to be deeply relevant to developers and grounded in their unique experiences (as opposed to a generic developer experience survey). To that end, we used much of what was gathered in the interviews to inform the survey, using interview reports of what caused a ``bad day'' to form questions (Likert scale and open-ended) about potential bad day scenarios.

Section 1 collected basic demographic information about participants, including their experience level, role, development type, location, work-from-home status, and gender identity. Figure \ref{fig:surveydemographics} shows the demographic spread of respondents of the survey. Interestingly, we had many more senior developers respond to the survey. The gender and role demographics are skewed towards male and individual contributors, but they do reflect the overall developer population.

\begin{figure}[ht]
    \centering
    \includegraphics[width=0.4\textwidth]{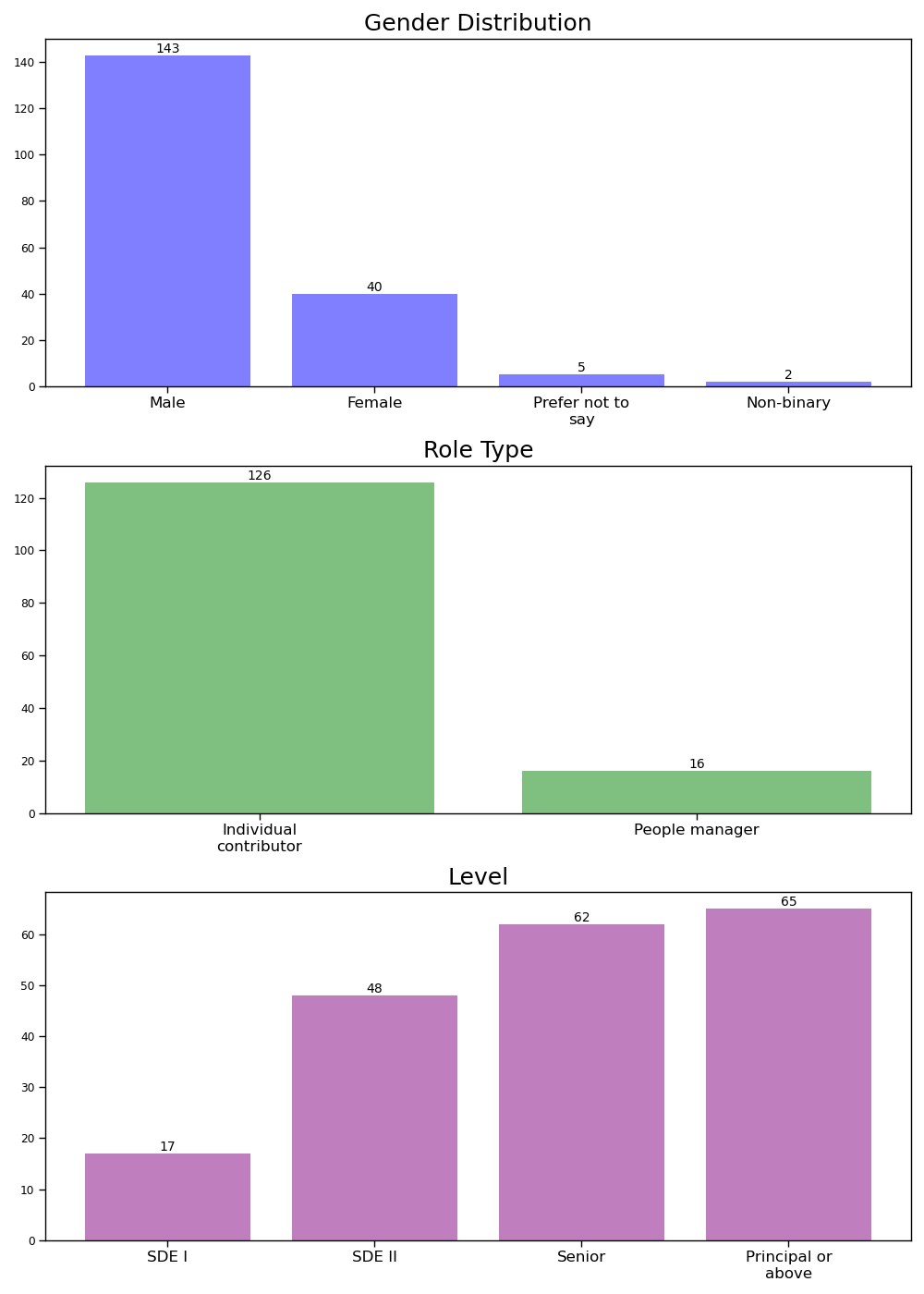}
    \caption{Demographic Profile of Survey Participants}
    \label{fig:surveydemographics}
    \vspace{-15pt}
\end{figure}

Section 2 focused on the perception of participants about the definition of what constitutes a ``bad day" as a developer, the frequency to which they are currently experiencing ``bad day'', and the perceived likelihood of various scenarios they encounter at work contributing to ``bad day'' presented as Likert scale questions. These Likert-style questions included queries about their rating of how likely blockers such as long build times, flaky tests, pull requests, on-call duties, meeting times, legacy code changes, and issues with engineering systems cause them to have a bad day. Here is a small example set of some questions from the survey:\\

\small
\begin{enumerate}
    
    \item In the last month, how many bad days have you had?

    \item How did a bad day impact your work and your personal life?

    \item In the following questions, you will be presented with various issues and scenarios related to your work. Please select how frequently each of the scenarios causes you to have a bad day. (Options were Always, Often, Sometimes, Rarely, Never, Not Applicable)
    \begin{enumerate}
        \item Long build time
        \item Your code was backed out, and it wasn't your fault
        \item Your code review is taking a long time
        \item Your VPN keeps dropping
        \item You had to debug into systems you don't own
        \item You have more than half your day filled with meetings
        \item You get negative feedback on a PR
        \item You are having a hard time getting support when having issues
    \end{enumerate}
\end{enumerate}

\normalsize

Section 3 offered participants the opportunity to opt in to participate in a daily short diary study. Participants who signed up would receive messages toward the end of every workday asking whether they experienced a ``bad day'' at work. If they responded "yes," they were asked a brief, open-text follow-up question about the contributing factor. 79 out of the 214 survey respondents participated in the daily diary study - a 36\% participation rate (Figure \ref{fig:datacollection}).

Section 4 sought participant consent to collect anonymized telemetry data related to their work, including check-ins, pull requests, code review time, meeting hours, and incident reports. The goal of the telemetry data was to allow the researchers to quantitatively measure how the baseline data for people who reported ``bad days'' differed from those people who did not report a ``bad day'' for the different specific issues listed in the second section of the survey. 131 out of the 214 survey respondents consented to have their telemetry data analyzed by our research team - a 61\% consent rate (Figure \ref{fig:datacollection}).

\subsubsection{Diary Study}

We conducted a diary study with 79 participants to capture daily data on the experiences of the developers, looking to uncover factors causing ``bad days'' at work. This longitudinal approach complemented the retrospective nature of the interview and survey and allowed us to observe how challenges and emotions fluctuate over time and in response to specific events or conditions that impact the productivity of developers within our organization. Specifically, the diary study allowed us to see what \textit{actually} caused a ``bad day'', in the moment, as opposed to what people \textit{believed} would cause a ``bad days''.

As described in the previous "Survey" section, participants for the diary study were recruited through a question at the end of the survey asking them to sign up. Those who consented to participate were prompted once every day during the workweek to visit an online questionnaire that asked about their experience for that day. The first question asked them to rate their day as either good, neutral, or bad. If the participants selected either good or neutral, the survey would end. However, if they selected bad, a freeform textbox was activated, encouraging them to share their experience of what made them have a ``bad day''. Overall, through this approach, we obtained longitudinal data on factors causing ``bad days'' for developers and any associated information the developer shares about them.

\subsubsection{Telemetry Study}

We collected telemetry and system data for developers who consented to investigate the correlation between poor system experience and reports of bad developer days. Specifically, the telemetry analysis focused on the measurable factors listed from the interviews and surveys and sought to validate the bad day factors identified through actual system data. This analysis focused mainly on the pull request and build times telemetry data. To protect the privacy of the developers, all telemetry data was anonymized and only accessible to members of the research team. We only collected metadata about development outcomes, such as how much time it took to approve their pull requests, the number of people that participated in the code review, and how long their build took to complete, among other related metadata. We provided the developers with a detailed explanation of the data being collected and informed them that they could opt out of telemetry collection at any time without affecting their employment or participation in the other aspects of the study.

\subsection{\textbf{Data Analysis}}

\paragraph{Interview Data} Data from the interviews was analyzed using a thematic analysis process that involved three steps. First, we immersed ourselves in the data by reading and re-reading the transcripts to better understand its contents and contexts. This allowed other members of the research team who were not part of the interview sessions to familiarize themselves with the data. Next, we commenced data analysis by conducting open coding of the interview transcripts to identify relevant concepts and patterns. Following the open coding, we transitioned to axial coding which allowed us to group initial codes together into potential themes and sub-themes. Next, we reviewed and refined the final themes generated from this process to ensure they accurately represented the data and then used the themes to conduct a thematic analysis that yielded the data we reported in this study.

\paragraph{Survey Data} We analyzed categorical data from the survey using descriptive statistics, correlation analysis, and cross-tab analysis. First, we conducted a descriptive statistics analysis to gain a general understanding of the data. We then followed up with a frequency distribution analysis to determine the prevalence of reported bad days and the common scenarios and blockers identified in the survey. To achieve this, we converted categorical responses into numerical values on a scale of 1 to 5 for each question/answer type as shown below:

\begin{itemize}
    \item Frequency of occurrence questions:\{``Always'': 5, ``Often'': 4, ``Sometimes'': 3, ``Rarely'': 2, ``Never'': 1, ``Not applicable'': 0\}
    \item Agreement level questions:\{``Strongly agree'': 5, ``Agree'': 4, ``Neither agree nor disagree'': 3, ``Disagree'': 2, ``Strongly disagree'': 1, ``Not applicable'': 0\}
    \item Likelihood questions:\{``Very likely'': 5, ``Somewhat likely'': 4, ``Neither likely nor unlikely'': 3, ``Somewhat unlikely'': 2, ``Very unlikely'': 1, ``Not applicable'': 0\}
\end{itemize} 

Next, we conducted a cross-tabulation analysis to examine how responses varied across different demographic groups, such as experience level, role, and location, revealing potential differences in the frequency of bad days using the Chi-square test for each pair of variables. These methods helped us identify any potential associations between, for example, experience level and the frequency of encountering specific blockers.

Finally, we conducted an open coding of survey responses to uncover the key themes causing bad days for developers.

\paragraph{Diary Data Analysis} Since the daily diary study only had 2 questions, we looked at descriptive statistics for the first question and then hand-coded the open-ended answers to find common themes that cause bad days for developers.

\paragraph{Telemetry Analysis} We collected six months of system and telemetry data of developers who signed up to investigate the potential relationships between telemetry data and reported bad days. The crucial aspect of this analysis involved linking survey responses with telemetry data through anonymized identifiers. Developers who reported that pull requests caused them to have a ``bad day'' were assigned to Group 1, while developers who reported that it did not cause them to have a ``bad day'' within the time frame of our study were assigned to Group 2. This approach allowed us to conduct a deeper investigation into how technical factors, such as lengthy build times or delayed pull requests, might contribute to reported bad days by conducting independent t-tests on the means of these metrics between Group 1 and Group 2 since our data is continuous (not categorical) and has a roughly normal distribution.

Overall, through these multiple research approaches, we were able to identify factors that cause ``bad days'' for developers, describe how they impact them, and validate some of the concerns of developers through telemetry analysis of two measurable factors that were identified from the interviews, survey, and diary studies. We report the findings from this analysis in detail in the following section.

\section{Findings}

\subsection{\textbf{RQ1: What factors cause a bad day for developers?}} We triangulated findings for this research question using the results from our interviews, survey, and diary study.

\subsubsection{\textbf{Interview findings}} Our interviews revealed three major themes that cause ``bad days'' for developers: tooling and infrastructure issues, process inefficiencies, and issues around team dynamics. Within those issues, deeper concerns were identified and are shared in the following sections.

\paragraph{Tooling and Infrastructure Issues} Participants complained consistently about how unreliable tools and infrastructure were a major source of frustration and a factor that caused ``bad days'' for them. Some of the issues identified included flaky builds and tests, issues around slow builds and deployments, outdated and clunky user interfaces, and generally unreliable or broken tools that block them from completing their tasks. For instance, one of the developers P04 mentioned, \textit{"the build tools, the environment tools, ADO, Pull requests, Git, Visual Studio automation, every single tool that we use feels like it's barely working most days,"} highlighting how the pervasiveness of poor tooling and engineering systems is disrupting their workflow. 

\paragraph{Process inefficiencies} Issues around unclear project ownership, lack of documentation and knowledge sharing, and rapidly changing team priorities were other common factors that caused ``bad days'' for developers. For instance, one of the developers mentioned that \textit{"the whole initiative [around documentation] was shut down, which is a huge bummer because documentation is definitely not a solved problem. And when you can't find the knowledge you need to get your work done, that's very frustrating."} Another developer mentioned process inefficiencies around item tracking commenting that \textit{"there's a remarkable amount of confusion about, like how work should be tracked, which is remarkable for, you know, our company is [very] old, and we're still unsure how to track work [leading to a cascade of inefficiencies.]"} This issue was more common among senior and principal developers.

\paragraph{Issues around Team dynamics and communication} Developers often complained about difficulties in collaboration, communication breakdowns, unresponsive team members, and interpersonal conflicts. Some developers highlighted how technical issues often lead to ``bad days'' that infect the overall team morale and often degrade trust levels. For instance, a developer (P03) remarked that \textit{"I'm sure some of that energy [from the frustration of facing technical issues] carries over and then impacts the team, you know like smiles are contagious and grumpiness is contagious."} Other developers highlighted how intra-team conflict caused them to have a ``bad day'' in the past and how they resolved it, saying \textit{"[I was having a bad time working with my prior team and solving] it was honestly, largely a matter of changing the, you know, the working group that I am collaborating with on a day to day basis."}  This issue was most common among junior developers.

\subsubsection{\textbf{Survey Results}} Due to the length and question types asked in the survey, the results are broken down into three sections: developers' definition of bad days, top factors causing bad days, and frequency of bad days (with analysis of demographics).

\paragraph{Developers Definition} As stated previously, our goal was not to prescribe a definition of ``bad days", rather, we wanted to hear what developers thought caused a bad day. Figure \ref{fig:baddayopencodingfrequency} shows the top 15 coded responses when developers were asked what a bad day means to them - the top 3 are Engineering System Friction, feeling Blocked or Stuck, and Poor Productivity. This is very similar to the findings from the Interviews.

\begin{figure}[ht]
    \centering
    \includegraphics[width=0.4\textwidth]{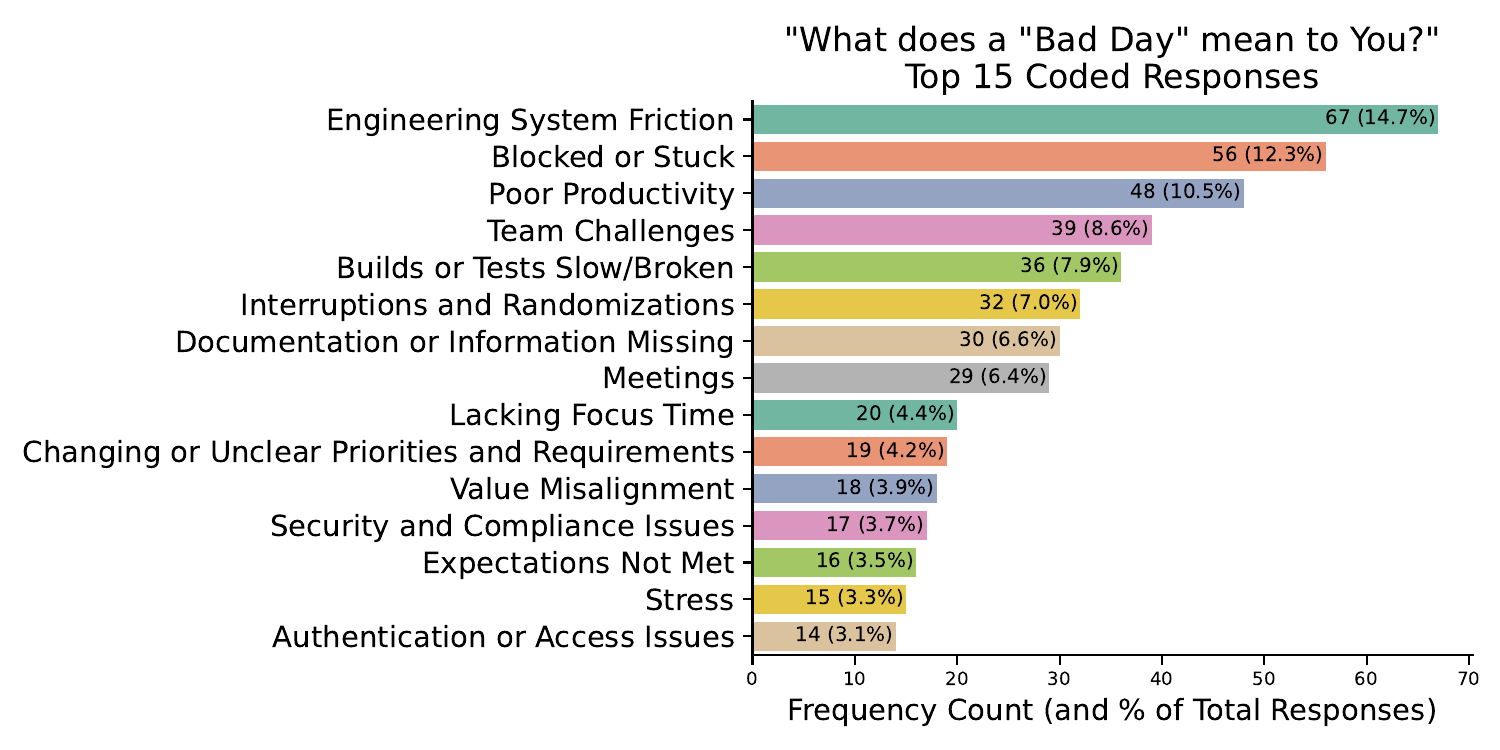}
    \caption{Manual coding of the open-ended question "What Does a ``Bad Day'' Mean to You?" revealed the most frequent responses.}
    \label{fig:baddayopencodingfrequency}
        \vspace{-7pt}
\end{figure}

\paragraph{Top Factors} From the Likert-style questions, we were able to rank factors causing developers to have bad days - Table \ref{table:topblockers} lists the top 15 with their mean scores. "PR delayed due to reasons beyond you and your team's control (flaky tests, transient issues, build failure, unresponsive reviews, etc.)" was the most common factor that causes a ``bad day'' for developers \textit{(mean score=4.22)}. In addition, many subjective factors like team dynamics or lack of support surfaced in the top 10 which is in line with what we observed in the interviews.

\begin{table*}[h!]
    \centering
    \caption{Blockers and Mean Scores}
    \begin{tabular}{|c|l|c|}
        \hline
        \textbf{Rank} & \textbf{Blocker} & \textbf{Mean Score} \\
        \hline
        1 & \parbox{10cm}{PR delayed due to reasons beyond you and your team's control (flaky tests, transient issues, build failure, unresponsive reviews, etc.)} & 4.22 \\ \hline
        2 & \parbox{10cm}{You feel like you didn't get anything done} & 4.20 \\ \hline
        3 & \parbox{10cm}{Your laptop is slow/laggy} & 4.15 \\ \hline
        4 & \parbox{10cm}{You are having a hard time getting support when having issues} & 4.11 \\ \hline
        5 & \parbox{10cm}{You have more than half your day filled with meetings} & 4.02 \\ \hline
        6 & \parbox{10cm}{People and teammates close to you affected by events like layoffs, covid, etc.} & 3.96 \\ \hline
        7 & \parbox{10cm}{Your VPN keeps dropping} & 3.90 \\ \hline
        8 & \parbox{10cm}{You had to fix something that wasn't your responsibility, but blocked you nonetheless} & 3.72 \\ \hline
        9 & \parbox{10cm}{You get assigned work you weren't expecting} & 3.68 \\ \hline
        10 & \parbox{10cm}{Teams isn't working} & 3.66 \\ \hline
        11 & \parbox{10cm}{You had to debug deep into systems you don't own} & 3.65 \\ \hline
        12 & \parbox{10cm}{Your manager shares tough feedback} & 3.52 \\ \hline
        13 & \parbox{10cm}{You need to access some kind of data you haven’t accessed before} & 3.52 \\ \hline
        14 & \parbox{10cm}{You had an issue with Torus} & 3.46 \\ \hline
        15 & \parbox{10cm}{Your teammates are grumpy/short with you} & 3.34 \\ \hline
        \end{tabular}%
    \label{table:topblockers}
\end{table*}

When we reviewed responses from different demographics, we found that SDE IIs and Senior developers report \textit{``having a hard time getting support when having issues''} as the most common factor that causes them to have bad days compared to other levels. Concerns around \textit{``slow/laggy laptop''} were distributed across different levels, but slightly more prevalent among SDE IIs and Senior Developers. In contrast, \textit{"you have more than half your day filled with meetings"} was a challenge most prevalent among Principals and Senior Developers. Furthermore, concerns around \textit{``slow/laggy laptop"} causing a ``bad day'' were more prevalent among those working from home. Similarly, individuals working primarily from home report a higher prevalence of meeting-heavy days. 

\paragraph{Frequency of ``Bad Days''} As shown in Figure \ref{fig:baddaysresponses}, 35\% of survey respondents say that they experienced a ``bad day" in the last month, and the distribution is roughly normal.

\begin{figure}[ht]
    \centering
    \includegraphics[width=0.47\textwidth]{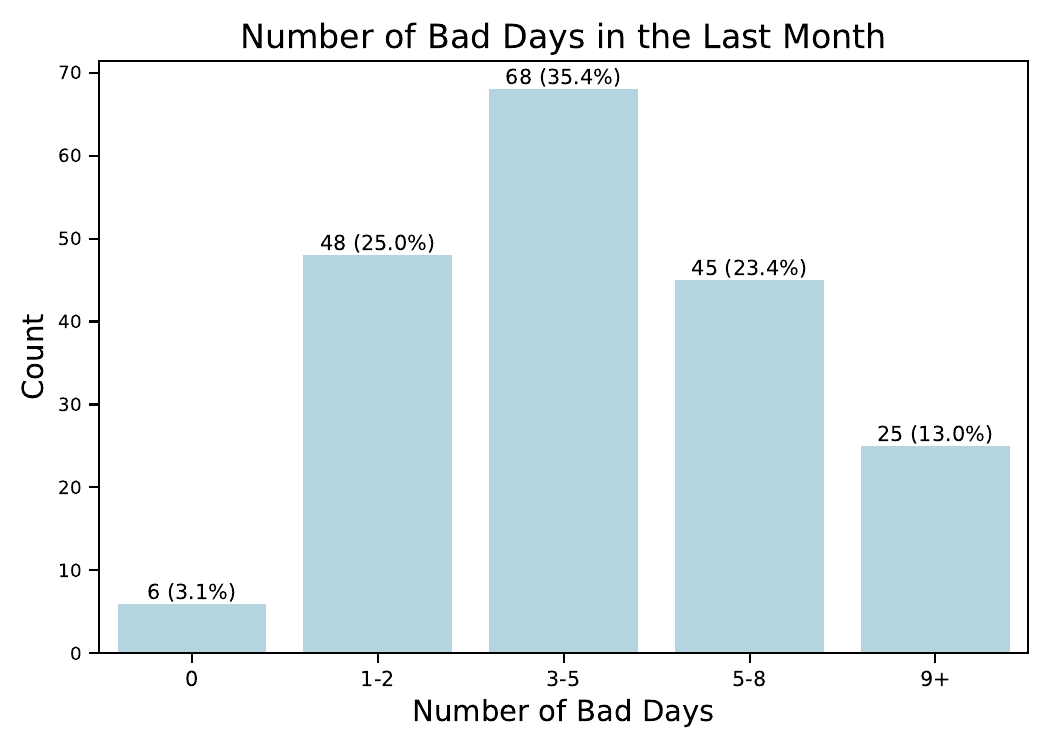}
    \caption{Self-Reported Frequency of Bad Days}
    \label{fig:baddaysresponses}
    \vspace{-15pt}
\end{figure}

We also did a cross-tabulation using a Chi-Square test between the frequency of ``bad days'' and demographic cohorts. We chose the Chi-Square test because it assesses whether there is a statistically significant relationship between two categorical variables. We did not find any statistically significant differences here which is a good thing since we do not want any demographic cohort to experience a higher prevalence of ``bad days'' than others.

\subsubsection{\textbf{Diary study results}} 79 unique respondents submitted 775 diaries over the duration of the study. Only 20\% of the responses said that the developer was having a ``bad day".


We also analyzed the freeform text response which was only submitted if a developer had a ``bad day''. This revealed the development environment infrastructure and technical issues were the most common challenges that caused developers to have a ``bad day'', very similar to the findings from the interview and survey. Interestingly, health or emotional challenges were a very close second. 



\begin{tcolorbox}[colback=gray!10, colframe=gray!40, title=Summary of Results from RQ1, coltitle=black, fonttitle=\bfseries, left=5pt, right=10pt]
Concerns around developer tooling/infrastructure, inefficient processes, and difficult interpersonal dynamics emerged as key factors 
causing bad days across the interview, survey, and diary methods.
\end{tcolorbox}

\subsubsection{\textbf{Discussion - RQ1}}

Findings from our analysis revealed a complex picture of developer experience within our organization and particularly showed that ``bad days'' are not simply a matter of technical challenges. These findings go against the typical expectation that improving technical infrastructure will enhance the developer experience. Instead, it aligns with results from prior research \cite{murphy2019predicts,storey2019towards} which has shown that both technical and non-technical issues impair developer experience, suggesting a need to pay closer attention to exploring how these themes interact with each other and examine ways of mitigating them through a holistic approach that focuses on the multifaceted nature of the issues.


\subsection{\textbf{RQ2: In what ways do developers describe how ``bad day'' factors impact them and their work?}}  Results from our interview and survey studies showed how developers reported that ``bad days'' negatively impact their work and well-being.

\subsubsection{\textbf{Interview findings}} Senior and junior developers described the impact of ``bad days'' on them in different ways. Specifically, our analysis showed that frustration, annoyance, and anger stemming from unexpected roadblocks and perceived inefficiency were the common ways senior developers described the impact of bad factors on their work. They described these feelings typically leave them exhausted due to long hours and the mental strain of dealing with ``bad days'', and a constant re-occurrence of ``bad days'' also leads them to disillusionment and cynicism from the perceived lack of action by the organization to mitigate these factors. For instance, while hinting at disillusionment, P04, a senior developer commented \textit{"And then eventually you're just like, why am I even driving on this road anymore?"}, signaling how frequent ``bad days'' cause them to be disillusioned and to consider looking for a new job. Another senior developer P02 also shared about their disillusionment saying, \textit{"Well, I mean when I've had a lot of consecutive ``bad days'' like that's when I start pulling up job boards and seeing, you know what other opportunities are there."} Another senior developer P16 also mentioned that \textit{"I would say on the on the worst of the ``bad days'' that I've had, those are the days where I actually just stopped working and go to LinkedIn and start looking at other jobs."}. Other than feeling the need to quit their jobs, developers mentioned feeling left behind and treated unfairly, like P02 saying, \textit{"I could tell that the impact it has on me is I'm less satisfied with my work in general. I feel like, well, you know, I can see all of these other people and they are chilling eating food. Like, I feel treated unfairly."}

Analysis from our interviews revealed the common impact of ``bad day'' factors on junior developers is that it leads to guilt and self-doubt because they attribute the cause of the problems to their own incompetence, instead of a systemic issue from the organization. For instance, a junior developer P12, mentioned that whenever they encounter such issues, \textit{"I start to think like, yeah, like, am I just not competent enough to fix this and I'm just getting super unlucky right now. It's all like pretty frustrating."}

\subsubsection{\textbf{Survey results}}

Findings from our analysis of the open-ended question on the impact of ``bad days'' in the survey revealed four key themes about how ``bad days'' impact developers, including concerns about reduced productivity and its impact on their career, self-doubt, and imposter syndrome arising from the frequent occurrence of bad days, extended work hours due to self-guilt, and overall increased stress and anxiety levels that spills over to their personal life. Figure \ref{fig:impactofbaddayscodedanswers} shows the frequency of these answers, and stress and health make up 3 out of the top 5 responses.

\begin{figure}[ht]
    \centering
    \includegraphics[width=0.47\textwidth]{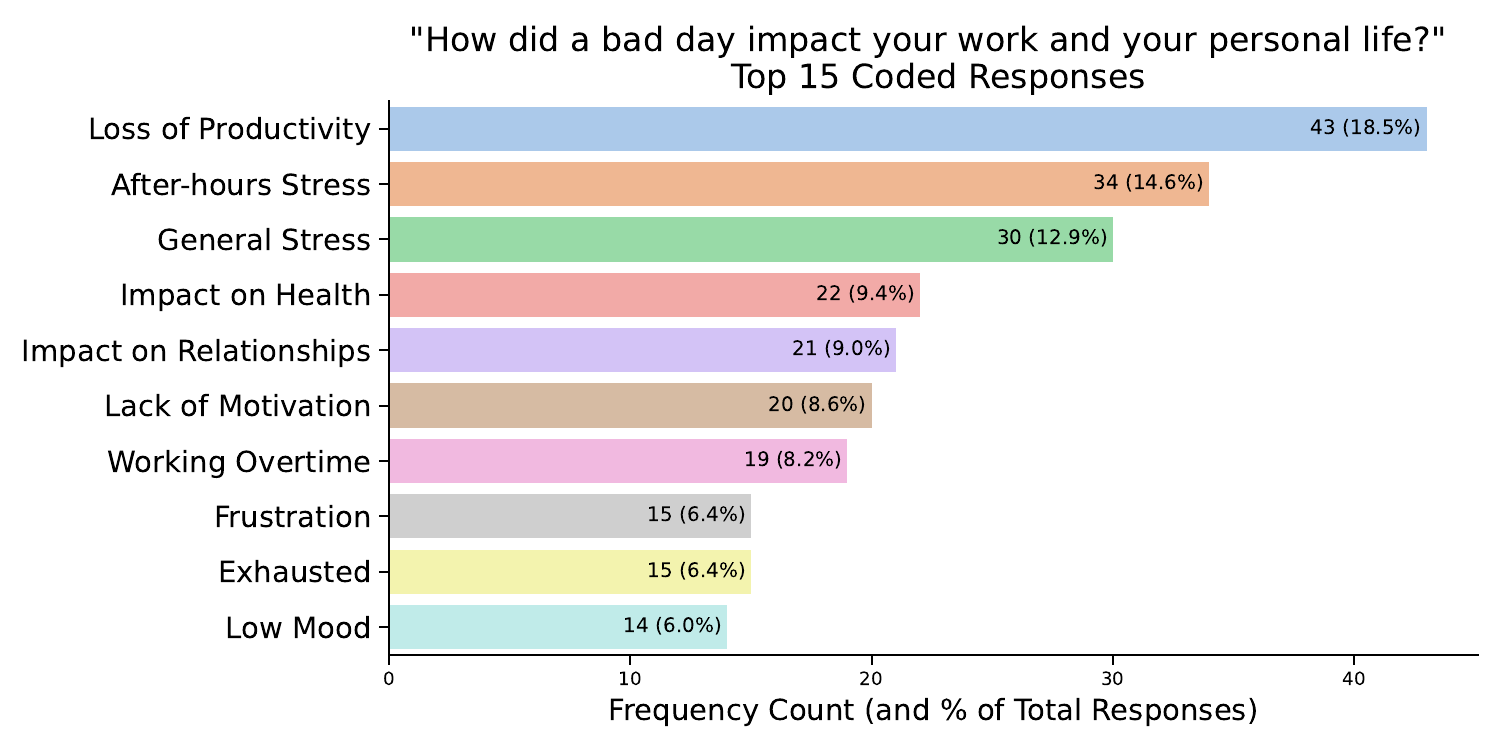}
    \caption{Coded Answers to "How did a Bad Day impact your
    \\ work and your personal life?"}
    \label{fig:impactofbaddayscodedanswers}
    \vspace{-7pt}
\end{figure}

Regarding concerns around reduced productivity and career implications, our analysis showed that ``bad days'' were consistently linked to decreased productivity, both in terms of quantity and quality of work. Specifically, developers reported feeling unable to focus, losing motivation, and struggling to complete tasks. For instance, a developer shared their experience about productivity, stress, and career, mentioning that \textit{" [a ``bad day''] makes progress difficult and feels like you are spinning wheels or wasting time. This creates anxiety and depression, as you fear for your job security due to lack of progress on goals, and leads to frustration for entire days at a time."} Another developer shared their experience about productivity loss as a result of a bad day mentioning that \textit{"a ``bad day'' impacts my work by reducing my productivity, causing delays in project timelines, and leading to a lower quality of work, such as missing edge cases."}

Following ``bad days'', developers often experience self-doubt and imposter syndrome, leading them to question their abilities and to feel like they are not meeting expectations. For instance, a developer sharing their experience following a ``bad day'' mentioned that \textit{"After a ``bad day,'' I begin to feel frustrated and guilty for not getting any work done and get hit by imposter syndrome. A ``bad day'' at work definitely carries on into my personal life after hours too. If the bad day is caused by unproductivity, I will think about the task/bug I'm working on while I'm off and feel stressed about getting it done."} Another developer mentioning the impact of bad days on their confidence remarked that \textit{"[a bad day] made me feel less confident in my abilities, resurfac[ing] imposter syndrome."}

Extended work hours due to self-guilt was another common impact of ``bad days'' on developers. This scenario typically arises as developers attempt to compensate for lost hours during the workday by resorting to working longer hours to catch up. This dynamic disrupts work-life balance leading to less time to rest, decompress, and recover from work. For instance, a developer sharing their experience about this dynamic mentioned that \textit{"[a ``bad day''] makes it difficult for me to stop work at 5 or 6 pm and spend time with my family because I feel like the fact that I didn't accomplish anything means I'm not "pulling my weight".  Therefore, I need to spend more time in the evening "catching up" so that I can justify my existence as an employee."} Another developer sharing a similar experience mentioned that \textit{"I have had to work after hours or on weekends to be able to focus and finish my work. This has resulted in not having enough personal time to spend on my kids extracurricular activities and sadly feel like I'm hurting their academic and future careers. I have also not been able to keep up with exercise or have time to prepare healthy meals and have had to rely on takeout."}

Overall, developers frequently reported that ``bad days'' led to increased levels of stress and anxiety which spills over into their personal lives, affecting sleep, relationships, and overall well-being. Other dimensions of the emotional impact of ``bad days'' on developers included reports of emotional exhaustion, feeling drained, frustrated, and overall in a negative mood. A developer sharing their experience remarked that \textit{"Often times I feel so emotionally exhausted after a day full of frustrating events that my evening is also essentially a write-off."} This highlights how ``bad days'' impact their ability to enjoy their personal time and engage in activities outside of work. Another developer sharing their experience mentioned that \textit{"I have trouble sleeping those days or keeping up with chores because I'm emotionally exhausted [after work]."}

\begin{tcolorbox}[colback=gray!10, colframe=gray!40, title=Summary of Results from RQ2, coltitle=black, fonttitle=\bfseries, left=5pt, right=10pt]
Disillusionment, diminished productivity, self-doubt, after-hours stress, overtime, personal relationships, concerns about career growth, and imposter syndrome were the most common impacts of ``bad days'' by developers.
\end{tcolorbox}

\subsubsection{\textbf{Discussion - RQ2}}

Findings from our analysis revealed that senior developers often experience frustration, annoyance, and anger stemming from unexpected blockers and inefficiencies, while junior developers often experience guilt and self-doubt. This finding provides a pathway to understand the kinds of support developers of different levels might need and the ways to best support them. For instance, junior developers could be supported through mentoring programs within the organization that are designed to increase their self-confidence, knowledge, and productivity, while senior developers could benefit from participating in co-creation activities where they play a role in designing the systems and processes that are used within the organization to ensure it works for them. In the survey, we asked respondents to share with us how what improvements could be made to reduce the number of ``bad days'' they experienced. 
Many respondents clearly articulated some concrete changes that could be made to reduce ``bad days'' such as fewer meetings, improved documentation, faster build times, changes in authentication, faster code reviews, and changes their managers could work by giving more autonomy to their developers. Some of the proposed changes were very detailed (and therefore proprietary) but showed that asking this question can be important to discover changes that could be made to reduce ``bad days.''

\subsection{\textbf{RQ3: Can we use telemetry data to validate factors that cause ``bad days'' for developers?}} Results from our telemetry analysis will be reported under two headings including pull request and build time telemetry data. 
We decided to focus on examining pull requests and build telemetry data because they were among the highest-ranking bad day factors from our survey results that are measurable using system data. For instance, other high-ranking factors like \textit{"feeling like you didn't get anything done"} are not easily measurable using system data.

\subsubsection{\textbf{Pull Request Time}} The process of editing and improving code involves sharing the changes in a pull request which will be reviewed by the author's teammates and other colleagues before merging them into the main project repository \cite{badampudi2023modern,mantyla2008types,sadowski2018modern,bosu2015characteristics}. The reviewers on the pull request review the code for logical completeness, code quality, and security and are a key part of the software engineering process. We employed metrics that cover each part of the pull request, as a result, our analysis was based on the mean comparison of the following metrics:
\begin{itemize}
    \item TotalPullRequestTimeHours: Total duration of pull request from open (non-draft state) to close in hours
    \item DwellTimeHours: Time from pull request open to pull request touched by a reviewer
    \item CodeReviewTimeHours: Time from pull request touched by a reviewer to pull request closed
    \item TotalReviewers: Number of reviewers on the pull request
\end{itemize}

Table \ref{tab:codereview_comparison} reports the findings from this statistical analysis. Group 1 indicated those who said Code Reviews cause them to have a ``bad day'', and Group 2 indicated those who said Code Reviews do not cause them to have a ``bad day''. We performed an independent t-test since our data is continuous (not categorical) and has a roughly normal distribution. 
We found two statistically significant results - Group 1 had 23.8\% (p-value $<$ 0.05) higher TotalPullRequestTimeHours and also 48.8\% higher DwellTimeHours (p-value $<$ 0.05). We did not find a significant difference in the code review time or number of reviewers.

\tiny
\begin{table*}[htbp]  
\caption{Comparison between those who said Code Reviews Cause a ``Bad Day'' [1] or Not [2]}
\centering
\renewcommand{\arraystretch}{1.5} 
\resizebox{\textwidth}{!}{ 
\begin{tabular}{|l|c|c|c|c|c|c|c|}
\hline
\textbf{Metric} & \textbf{Mean Group 1} & \textbf{Mean Group 2} & \textbf{Mean Diff (\%)} & \textbf{p-value} & \textbf{Significant?} & \textbf{Power} \\ 
\hline
TotalPullRequestTimeHours & 47.41 & 36.09 & 23.87 & 0.00 & True & 0.90 \\ 
\hline
DwellTimeHours & 22.49 & 11.51 & 48.84 & 0.00 & True & 1.00 \\ 
\hline
CodeReviewTimeHours & 18.09 & 21.12 & -16.77 & 0.22 & False & 0.23 \\ 
\hline
TotalReviewers & 2.43 & 2.38 & 2.00 & 0.34 & False & 0.16 \\ 
\hline
\end{tabular}}
\label{tab:codereview_comparison}
\end{table*}
\normalsize

\tiny
\begin{table*}[htbp]  
\caption{Comparison between those who said Long Builds Cause a ``Bad Day'' [1] or Not [2]}
\centering
\renewcommand{\arraystretch}{1.5} 
\resizebox{\textwidth}{!}{ 
\begin{tabular}{|l|c|c|c|c|c|c|c|}
\hline
\textbf{Metric} & \textbf{Mean Group 1} & \textbf{Mean Group 2} & \textbf{Mean Diff (\%)} & \textbf{p-value} & \textbf{Significant?} & \textbf{Power} \\ 
\hline
TotalBuildTimeMinutes & 49.74 & 36.65 & 26.32 & 0.00 & True & 1.00 \\ 
\hline
CoreBuildTimeMinutes & 47.08 & 33.89 & 28.02 & 0.00 & True & 1.00 \\ 
\hline
NonCoreBuildTimeMinutes & 13.67 & 11.35 & 17.00 & 0.07 & False & 0.40 \\ 
\hline
BuildSuccessRate & 89.21\% & 92.54\% & -3.60 & 0.15 & False & 0.60 \\ 
\hline
\end{tabular}}
\label{tab:codebuild_comparison}
\end{table*}
\normalsize

\subsubsection{\textbf{Build Process Time}} The build process measures the duration required to compile, test, and package code into a deployable format \cite{smith2011software,phillips2014understanding,yu2003removing,tu2001build}. Since the product is large and build time is dependent upon the availability of cache binaries from prior builds, a developer may encounter an especially long build time, even if they made no changes, due to none of the dependent binaries being cached. We report the results for the build process telemetry analysis based on the mean comparison of the following metrics:
\begin{itemize}
    \item TotalBuildTimeMinutes: Total duration of the build in minutes
    \item CoreBuildTimeMinutes: Total duration of the core aspects of the build in minutes
    \item NonCoreBuildTimeMinutes: Total duration of the non-core aspects of the build in minutes
    \item BuildSuccessRate: Number of successful builds out of total number of builds
\end{itemize}

Table \ref{tab:codebuild_comparison} reports the findings from this analysis in tabular format. Group 1 indicates those who said Long Builds cause them to have a ``bad day,'' and Group 2 indicates those who said Long Builds do not cause them to have a ``bad day.'' Once again, we performed an independent t-test since our data is continuous (not categorical) and has a roughly normal distribution.
Group 1 had a significantly longer mean build time (49.74 mins) compared to Group 2 (36.65 mins), with a 26.32\% difference (p-value $<$ 0.05).

In summary, Group 1 had statistically significant longer times for total pull request time, pull reuqest dwell time, and overall build time. Therefore, we can conclude that the concerns expressed by members of Group 1 are verifiable by their telemetry.

\begin{tcolorbox}[colback=gray!10, colframe=gray!40, title=Summary of Results from RQ3, coltitle=black, fonttitle=\bfseries, left=5pt, right=10pt]
Telemetry data analysis validated that there is a statistically significant difference between the pull request and build process log data of those who reported that pull requests and build process caused them to have bad days within the study time frame versus those who reported that those factors did not cause them to have a bad day.
\end{tcolorbox}

\subsubsection{\textbf{Discussion - RQ3}}

The telemetry data analysis provided empirical validation for the self-reported ``bad day'' factors. This validation is a critical step toward building data-driven approaches for assessing and improving developer experience. For instance, leveraging telemetry data, an organization can develop predictive models that can anticipate ``bad days'' and enable proactive interventions. Overall, while this approach has its limitations - such as the potential influence of other variables not captured in the telemetry data - it demonstrates the value of combining qualitative and quantitative data for a better understanding of factors that cause ``bad days'' for developers and impede developer productivity and experience.

\section{Limitations}
There are a few limitations that might impact the reproduction of this study. Our sample population was limited to a sub-population within our organization. Thus, it is possible that sampling a different population will yield a different ranking of factors that cause developers to have a ``bad day.'' Therefore, other organizations seeking to replicate this study should focus on our process of eliciting the ``bad day'' factors and less on the result since the outcomes will most likely differ between organizations.

In addition, participants who responded to calls to participate in the interviews, survey, and diary studies might be those who have experienced more ``bad days'' in the past, hence data collected from our engagement with them might not represent the views of most developers within the organization. Different cultures also have varying levels of comfort in expressing their views in a stronger or weaker manner which may influence the responses as well. Again, organizations that replicate this study should focus on the research methods outlined here rather than the specific results.

    \vspace{-5pt}

\section{Conclusion \& Implication}
Findings from our research revealed a variety of factors, including technical factors like frequent auth prompts, slow build times, and long pull request times, and non-technical factors like too many meetings, difficulty finding help when blocked, and intra-team conflicts that cause developers to have ``bad days'' at work. Most importantly, our research revealed a vicious cycle arising from ``bad days'' which warrants the attention of scholars and further research. For instance, findings from our analysis revealed that ``bad days'' often cause sleep issues for developers when they get home and we know from prior research that cognitive fatigue affects performance which means that when they get home and do not get enough rest, they come back to work the next day are tired - making them feel unproductive and it becomes a cycle. This theme needs further attention from scholars as the focus is often directed towards looking at tool friction in isolation as a factor that causes ``bad days'' for developers, but rarely do we take a holistic look through the lens of developers of how their experience with these frictions and blockers impact both their professional and personal life in ways that leave them both unproductive and potentially, cognitively and physically drained.

%
\section*{Acknowledgements}
We thank all participants who completed an interview, survey or joined the diary study for helping us shape and complete the research.
We would also like to thank \textbf{Tim Bozarth}, the originator of Bad Developer Days at Microsoft, for his insightful feedback throughout all stages of this project. Thank you \textbf{Josh Pollock, David Speirs, and Barrett Amos} for their constructive and insightful feedback throughout all stages of this project.

\typeout{}
\bibliography{main}
\bibliographystyle{IEEEtran}

\end{document}